\documentclass[aps,prd,onecolumn,nofootinbib]{revtex4-2}

\usepackage{amsmath,amssymb}
\usepackage{graphicx}
\usepackage{hyperref}

\begin{document}

\title{Complementary Roles of Distance and Growth Probes in Testing Time-Varying Dark Energy}

\author{Seokcheon Lee}
\affiliation{Department of Physics, Sungkyunkwan University, Suwon 16419, Korea}

\date{\today}

\begin{abstract}
Distance measurements have long provided the primary observational constraints on the expansion history of the Universe and the properties of dark energy. However, because such observables depend on cumulative line-of-sight integrals over the Hubble rate, their sensitivity to time-dependent features of the dark energy equation of state is intrinsically limited. In this work, we examine this limitation from an information-based perspective using the eigenvalue structure of the Fisher information matrix constructed from distance, expansion rate, and growth observables. We show that distance and expansion-rate data generically produce a strongly hierarchical Fisher spectrum dominated by a single information mode, reflecting an irreducible loss of sensitivity to temporal variations in dark energy. This behavior can be traced directly to the integrated kernel structure of geometric observables. Growth measurements, by contrast, respond through differential dynamics and can introduce additional independent information directions. Using both controlled mock data and survey-like configurations representative of next-generation experiments, we find that the impact of growth information depends not only on its nominal precision but also on the structure of the data covariance. In simplified mock setups, growth measurements can partially activate a second information direction even at moderate precision. In Euclid-like configurations, however, the information remains effectively one-dimensional until growth precision reaches the percent level, below which a second mode emerges rapidly. These results clarify the complementary roles of distance and growth probes and provide a model-independent criterion for assessing the physical content of cosmological constraints on dynamical dark energy.
\end{abstract}

\maketitle

\section{Introduction}
\label{sec:Intro}

Over the past two decades, distance-based cosmological observations have played a central role in establishing the accelerated expansion of the Universe. Measurements of Type~Ia supernovae, baryon acoustic oscillations, and related distance indicators have provided remarkably consistent constraints on the background expansion history and have formed the empirical foundation of the standard cosmological model. These successes have been extensively discussed in the literature and reviewed in numerous observational and theoretical studies~\cite{SupernovaSearchTeam:1998fmf,SupernovaCosmologyProject:1998vns,SNLS:2005qlf,SupernovaCosmologyProject:2008ojh,SNLS:2011cra,SDSS:2014iwm,Planck:2018vyg,DESI:2025fxa}.

Beyond the determination of the present-day acceleration, a major goal of modern cosmology is to test whether dark energy (DE) exhibits time-dependent behavior. This question has motivated a wide range of phenomenological parametrizations of the DE equation of state (EoS), including the widely used CPL form and its variants, as well as non-parametric reconstructions
\cite{Chevallier:2000qy,Linder:2002et,Huterer:2000mj,Huterer:2002hy,Wang:2008zh,Zhao:2017cud}. However, it has also long been recognized that distance observables alone provide limited sensitivity to time variation in the DE EoS. This limitation has been discussed from multiple perspectives, including integral degeneracies, smoothing effects, and strong parameter correlations~\cite{Cooray:1999da,Maor:2000jy,Maor:2001ku,Bassett:2004wz,Huterer:2004ch,Clarkson:2007bc}.

The origin of this limited sensitivity can be traced to the cumulative nature of distance measurements. Luminosity and angular-diameter distances depend on integrals over the expansion history, which act to smooth localized or rapidly varying features in the underlying EoS~\cite{Lee:2025jrr}. As a result, even high-precision distance data tend to constrain only a restricted subset of possible DE evolution histories, often leading to strong degeneracies among parameters describing time variation~\cite{Crittenden:2005wj,Kunz:2007rk,Clarkson:2007pz,Sahni:2014ooa}.

In contrast, observables related to the growth of cosmic structure probe the expansion history in a qualitatively different manner.
Measurements of redshift-space distortions, weak gravitational lensing, and galaxy clustering provide access to the growth rate of matter perturbations, which responds more locally to changes in the expansion rate and gravitational dynamics~\cite{Kaiser:1987qv,Peacock:1996ci,Guzzo:2008ac,Song:2008qt,Reyes:2010tr,Beutler:2012px,BOSS:2016wmc}. As a consequence, growth observables have long been expected to offer complementary information on DE and gravity, particularly in the context of testing departures from a cosmological constant~\cite{Linder:2005in,Amendola:2007rr,Jain:2010ka,Weinberg:2013agg}.

A number of previous studies have explored the combination of distance and growth data, demonstrating improved constraints on DE parameters and modified gravity models when both classes of observables are used~\cite{Huterer:2006mva,Simpson:2012ra,DiValentino:2017iww}. Nevertheless, a clear and intuitive explanation of \emph{why} growth measurements are intrinsically more sensitive to time variation than distance probes has remained somewhat obscured, often buried in detailed parameter forecasts or model-specific analyses.

In this work, we revisit this issue from a unified phenomenological perspective. Rather than focusing on specific DE parametrizations, we analyze the structure of the Fisher information matrix associated with distance, expansion rate, and growth observables. By examining the eigenvalue spectrum of the Fisher matrix, we assess how many independent modes of DE evolution can be constrained by different classes of observations. We show that distance and expansion data generically produce a strongly hierarchical Fisher spectrum, dominated by a single eigenmode, reflecting the intrinsic degeneracy induced by their cumulative kernel structure. In contrast, the inclusion of growth measurements modifies the Fisher spectrum by lifting subdominant modes, thereby enabling sensitivity to the time dependence of the DE EoS.

Our analysis is directly relevant to current and upcoming surveys such as Euclid and LSST, which combine high-precision distance measurements with multiple probes of structure growth~\cite{LSSTScience:2009jmu,EUCLID:2011zbd,DESI:2016fyo}.
By clarifying the complementary roles of distance and growth observables, this work provides a transparent framework for interpreting future constraints on dynamical dark energy (DDE) and for assessing the observational requirements necessary to robustly test time-dependent cosmic acceleration.

The structure of this paper is as follows. In Sec.~\ref{sec:DisExPGro}, we introduce the distance, expansion rate, and growth observables considered in this work, emphasizing the distinct kernel structures that govern their response to DE evolution. We then describe the construction of the Fisher information matrix and the use of its eigenvalue spectrum as a diagnostic of parameter sensitivity in Sec.~\ref{sec:Fish}. In Sec.~\ref{sec:Hierarchy}, we demonstrate that distance and expansion data generically lead to a strongly hierarchical Fisher spectrum, effectively constraining only a limited subset of DE evolution modes. The impact of including growth measurements is analyzed in Sec.~\ref{sec:Growth}, where we show how additional eigenmodes are activated and sensitivity to the time dependence of the DE EoS is enhanced. In Sec.~\ref{sec:Implications}, we discuss the implications of these results for current and forthcoming surveys, with particular emphasis on Euclid-like configuration representative of next-generation experiments. Finally, we summarize our findings and outline their broader implications for testing DDE in Secs.~\ref{sec:Dis} and~\ref{sec:Con}.

\section{Distance, Expansion, and Growth Observables}
\label{sec:DisExPGro}

In this section, we summarize the cosmological observables considered in this work and emphasize the differences in how they respond to the underlying evolution of DE. Our focus is on the exact relations between the observables and the background expansion and growth histories, rather than on approximate parametrizations or survey-specific details.

\subsection{Distance and Expansion Kernels}

In a spatially flat, homogeneous, and isotropic Universe described by the Friedmann--Lemaître--Robertson--Walker metric, distance observables are determined by integrals over the Hubble expansion rate. The comoving radial distance $\chi(z)$ is given by
\begin{equation}
\chi(z) = \int_{0}^{z} \frac{c \, dz'}{H(z')}, \label{eq:comoving_distance}
\end{equation}
where $c$ is the speed of light in vacuum and $H(z)$ is the Hubble expansion rate. The luminosity distance $D_L(z)$ and angular-diameter distance $D_A(z)$ are then related to $\chi(z)$ through
\begin{equation}
D_L(z) = (1+z)\,\chi(z), \qquad D_A(z) = \frac{\chi(z)}{1+z},
\label{eq:distance_definitions}
\end{equation}
independent of any assumptions about the DE EoS beyond its contribution to $H(z)$. The expansion rate itself is determined by the Friedmann equation
\begin{equation}
H^2(z) = \frac{8\pi G}{3} \left[ \rho_m(z) + \rho_r(z) + \rho_{\rm DE}(z) \right] = H_0^2 \left[ \Omega_{m0} (1+z)^3 + \Omega_{r0} (1+z)^4 + \Omega_{\rm DE0} e^{S(z)} \right] , \label{eq:friedmann}
\end{equation}
where $\rho_{\rm DE}(z)$ evolves according to
\begin{equation}
\rho_{\rm DE}(z) = \rho_{\rm DE0} \exp\!\left[ 3 \int_{0}^{z} \frac{1 + \omega(z')}{1+z'} \, dz' \right] \equiv \rho_{\rm DE0} e^{S(z)}.
\label{eq:rho_de_exact}
\end{equation}
Here $H_0$ denotes the Hubble constant at the present epoch, $\Omega_{m0}$, $\Omega_{r0}$, and $\Omega_{\rm DE0}$ are the present-day density parameters of matter, radiation, and DE, respectively, and $\omega(z)$ is the DE EoS.

Equations~\eqref{eq:comoving_distance}--\eqref{eq:rho_de_exact} show explicitly that distance observables depend on cumulative integrals over the expansion history. As a result, variations in the DE EoS at different redshifts can compensate each other in the integrated quantities, leading to strong degeneracies when distance and expansion data are interpreted in terms of time-dependent DE models.

\subsection{Growth Observables}

The growth of matter density perturbations provides a complementary probe of cosmic acceleration. In the linear regime, the linear growth factor $D_{+}(a)$ satisfies the exact second-order differential equation
\begin{equation}
\frac{d^2 D_{+}}{d(\ln a)^2} + \left( 2 + \frac{d\ln H}{d\ln a} \right) \frac{dD_{+}}{d\ln a} - \frac{3}{2}\,\Omega_m(a)\,D_{+} = 0, \label{eq:growth_equation} 
\end{equation}
where $a=(1+z)^{-1}$ is the scale factor.  $\Omega_m(a)$ is the matter density parameter as a function of the scale factor, defined exactly as 
\begin{equation} \Omega_m(a) \equiv \frac{\Omega_{m0} H_0^2}{a^3 H^2(a)} .\end{equation}
Observationally, growth measurements are commonly reported in terms of the quantity $f\sigma_8(z)$, where the growth rate
\begin{equation}
f(z) = \frac{d\ln D_{+}}{d\ln a} \label{eq:growth_rate}
\end{equation}
is evaluated at the redshift of observation. Because Eq.~\eqref{eq:growth_equation} is a differential equation, changes in the expansion history induced by a time-dependent DE EoS affect the growth rate in a manner that is less strongly averaged over redshift than in distance observables.

The essential difference between distance and growth probes therefore lies in the mathematical structure of their response to DE evolution. While distance and expansion measurements encode information through integrals over $H(z)$, growth observables respond through the solution of a differential equation. As we demonstrate in the following sections, this distinction is directly reflected in the structure of the Fisher information matrix and determines how many independent modes of DE evolution can be constrained by a given set of observations.

\section{Fisher Matrix and Eigenmode Diagnostics}
\label{sec:Fish}

In this section, we introduce the Fisher information matrix used throughout this work and explain how its eigenvalue spectrum provides a transparent diagnostic of parameter sensitivity.  Our emphasis is on exact expressions and their physical interpretation, rather than on survey-specific details or approximate forecasting formulas.

\subsection{Fisher Matrix Construction}

Let $\boldsymbol{\theta} = (\theta_1,\theta_2,\ldots,\theta_N)$ denote a set of cosmological parameters describing the expansion history and the DE EoS.  In this work, we consider the parameter vector $\boldsymbol{\theta} = (\Omega_{m0}, H_0, \omega_0, \omega_a, \sigma_8)$, where $\omega_0$ and $\omega_a$ parametrize the time dependence of the DE EoS and $\sigma_8$ denotes the present-day rms amplitude of matter fluctuations on $8\,h^{-1}\,\mathrm{Mpc}$ scales. We consider a set of observables $\mathcal{O}_\alpha(z)$, which may include distance measures ($D_{L(A)}(z)$), expansion-rate data $H(z)$, and growth observables ($f \sigma_8(z)$) evaluated at a set of redshifts.

Assuming Gaussian-distributed measurement uncertainties, the Fisher information matrix is defined exactly as
\begin{equation}
F_{ij} = \sum_{\alpha,\beta} \frac{\partial \mathcal{O}_\alpha}{\partial \theta_i} \, \left( C^{-1} \right)_{\alpha\beta} \, \frac{\partial \mathcal{O}_\beta}{\partial \theta_j}, \label{eq:fisher_definition}
\end{equation}
where $C_{\alpha\beta}$ is the covariance matrix of the observables and all derivatives are evaluated at the fiducial cosmological model. In this work, the derivatives $\partial \mathcal{O}_\alpha / \partial \theta_i$ are computed exactly from the underlying relations between the observables and the background expansion and growth histories. For distance observables, these derivatives involve integrals over the Hubble expansion rate through Eqs.~\eqref{eq:comoving_distance}--\eqref{eq:rho_de_exact}, while for growth observables they are obtained by differentiating the solution of the growth equation~\eqref{eq:growth_equation} with respect to the parameters. No approximations are introduced at this stage beyond the assumption of linear perturbation theory for the growth of structure.

The Fisher matrix encodes the local curvature of the likelihood surface in parameter space. However, the interpretation of individual parameter uncertainties can be obscured by strong correlations among parameters, particularly when the observables depend on cumulative kernels. To expose the underlying structure of parameter sensitivity, it is therefore useful to examine the eigenmodes of the Fisher matrix.

\subsection{Eigenvalue Spectrum}

The Fisher matrix $F_{ij}$ is a real, symmetric, and positive semi-definite matrix. It can be diagonalized as
\begin{equation}
F = W^{\mathsf{T}} \, \Lambda_F \, W,
\label{eq:fisher_diagonalization}
\end{equation}
where $W$ is an orthogonal matrix whose rows define linear combinations of the original parameters, and
\begin{equation}
\Lambda_F = \mathrm{diag}(\lambda_1,\lambda_2,\ldots,\lambda_N)
\end{equation}
is the diagonal matrix of non-negative eigenvalues, ordered such that $\lambda_1 \ge \lambda_2 \ge \cdots \ge \lambda_N$.

Each eigenvalue $\lambda_n$ quantifies the amount of information carried by the corresponding eigenmode, defined as a specific linear combination of the parameters. Large eigenvalues correspond to well-constrained modes, while small eigenvalues indicate directions in parameter space that are weakly constrained or effectively unconstrained by the data. Therefore, the eigenvalue spectrum provides a basis-independent measure of how many independent combinations of parameters can be constrained by a given set of observables. A strongly hierarchical spectrum, characterized by $\lambda_1 \gg \lambda_2$, indicates that the data effectively constrain only a single dominant mode. In contrast, a flatter spectrum signals the presence of multiple independent modes with comparable constraining power.

This eigenmode-based perspective is particularly useful for understanding the limitations of distance and expansion observables. As shown in the following section, the cumulative kernel structure of these observables generically leads to a Fisher matrix with a pronounced eigenvalue hierarchy. The inclusion of growth measurements modifies this structure by introducing additional sensitivity through the differential nature of the growth equation, thereby lifting subdominant eigenvalues and increasing the number of independently constrained modes. In the remainder of this work, we use the eigenvalue spectrum of the Fisher matrix as a diagnostic tool to quantify and compare the constraining power of different combinations of cosmological observables.

\subsection{Numerical derivatives and implementation details}
\label{subsec:numerical}

The Fisher matrix elements are evaluated using numerical derivatives of the observables with respect to the cosmological parameters around a fixed fiducial model. Throughout this work, we employ symmetric finite-difference derivatives and explicitly verify the numerical stability of the resulting Fisher eigenvalue spectra. All numerical derivatives are evaluated around a standard fiducial $\Lambda$CDM cosmology, which serves as the reference point for the local linearization implicit in the Fisher-matrix formalism. Specifically, we adopt
\begin{align}
\Omega_{m0} = 0.30, \qquad H_0 = 70\ \mathrm{km\,s^{-1}\,Mpc^{-1}}, \qquad (\omega_0,\,\omega_a) = (-1,\,0), \qquad
\sigma_8 = 0.8,
\end{align}
corresponding to the cosmological-constant limit of the CPL parametrization. This choice is not intended to assume the absence of DDE, but rather to provide a neutral and well-defined expansion point for assessing whether time-dependent features of the DE EoS are observationally accessible in the first place. We have explicitly verified that the qualitative structure of the Fisher eigenvalue spectrum is insensitive to moderate shifts of the fiducial model within the observationally allowed parameter range.

For distance and expansion observables, parameter variations are propagated through the background evolution by recomputing $H(z)$ and the associated distance integrals for each shifted parameter set. All integrals are evaluated using adaptive quadrature routines, and the finite-difference step sizes are chosen such that the induced fractional changes in the observables remain well within the linear response regime. We have checked that the qualitative hierarchy of Fisher eigenvalues is insensitive to moderate variations of these numerical step sizes.

Growth observables require a distinct numerical treatment. For each parameter variation, the linear growth equation, Eq.~\eqref{eq:growth_equation}, is solved numerically using a standard Runge--Kutta integrator. The growth factor is normalized by imposing $D_{+}(z=0)=1$, and the growth rate $f(z)$ and $f\sigma_8(z)$ are constructed directly from the numerical solution and its logarithmic derivative. This procedure ensures a consistent propagation of parameter variations through both the background expansion and the perturbation dynamics.

Two classes of data configurations are considered in the numerical analysis: controlled mock data experiments and an Euclid-like setup. The theoretical modeling, fiducial cosmology, and numerical differentiation scheme are identical in the two cases. They differ only in their redshift sampling, error models, and covariance assumptions, which are summarized explicitly in Table~\ref{tab:mock_vs_euclid_numeric}.

Before proceeding, we emphasize the numerical nature of the Fisher construction used throughout this work. All Fisher matrices are computed deterministically from finite-difference derivatives of theoretical observables evaluated at a fixed fiducial cosmology. No mock realizations or Monte Carlo sampling are employed. The numerical setup is intentionally minimal: our goal is not parameter forecasting, but to isolate the intrinsic information structure encoded in different classes of observables.

\subsection{Methodology independence of the information geometry}
\label{sec:methodology_independence}

Although different inference techniques---including $\chi^2$ minimization, Bayesian posterior analyses, principal-component approaches (PCA), and Gaussian-process (GP) reconstructions---employ distinct statistical methodologies, they all probe the same local information geometry of the likelihood in the vicinity of its maximum. In this regime, the curvature of the log-likelihood defines the Fisher information matrix, which governs the response of the observables to variations in the model parameters \cite{Tegmark:1996bz,Trotta:2008qt}.

As a result, differences between inference methods primarily reflect how this local information is explored or regularized, rather than differences in the information content itself. In particular, PCA explicitly diagonalizes the Fisher matrix, while Bayesian constraints reduce to the same structure under a quadratic (Laplace) approximation, to leading order and within the local Gaussian approximation~\cite{MacKay:2003,Trotta:2008qt}. GP reconstructions, although formally non-parametric, can be viewed as inducing an effective Fisher-like geometry through kernel-weighted covariances that control the number of recoverable functional degrees of freedom~\cite{Seikel:2012uu,Rasmussen:2006}.

The strongly anisotropic eigenvalue spectrum found for distance observables reflects a structural property of the observable--theory mapping, rather than a limitation specific to any individual inference technique. Improving the precision of geometric measurements primarily tightens constraints along the dominant Fisher direction, without activating additional independent modes. A concise comparison of these inference methodologies and their relationship to the Fisher information matrix is summarized in Table~\ref{tab:methodology_fisher}.

\begin{table*}[t]
\centering
\caption{Comparison of different inference methodologies and their connection to the Fisher information matrix. Although these approaches employ distinct statistical frameworks, the locally accessible information is governed by the same Fisher geometry of the likelihood.}
\label{tab:methodology_fisher}
\begin{ruledtabular}
\begin{tabular}{llll}
Methodology
& Inference object
& Fisher-matrix connection
& Representative references \\
\hline
$\chi^2$ minimization
& Best-fit parameters
& Curvature of $\chi^2$ at minimum
& \cite{Tegmark:1996bz} \\

Bayesian posterior
& Posterior distribution
& Quadratic (Laplace) expansion of $\ln \mathcal{L}$
& \cite{MacKay:2003,Trotta:2008qt} \\

PCA
& Independent parameter modes
& Eigen-decomposition of Fisher matrix
& \cite{Huterer:2002hy,Huterer:2004ch} \\

GP
& Function reconstruction
& Kernel-weighted effective Fisher covariance
& \cite{Seikel:2012uu,Rasmussen:2006} \\
\end{tabular}
\end{ruledtabular}
\end{table*}

\section{Hierarchy of Distance Constraints}
\label{sec:Hierarchy}

In this section, we examine the structure of the Fisher information matrix obtained from distance and expansion observables alone and demonstrate that it generically exhibits a strong hierarchy in its eigenvalue spectrum. This hierarchy can be traced directly to the causal and cumulative structure of the distance kernels, which encode information only from earlier epochs.
In the following, we use $\chi(z)$ to denote the comoving radial distance and use $D(z)$ as a generic notation for distance observables such as $D_L(z)$ and $D_A(z)$, which are directly related to $\chi(z)$ through Eq.~\eqref{eq:distance_definitions}.

\subsection{Distance-only Fisher matrix}

We begin by considering a Fisher analysis based solely on distance and expansion observables, including luminosity and angular-diameter distances and direct measurements of the Hubble expansion rate. From Eqs.~\eqref{eq:comoving_distance}--\eqref{eq:rho_de_exact}, the response of a generic distance observable $D(z)$ to a perturbation $\delta \omega(z)$ in the DE EoS can be written exactly as
\begin{equation}
\delta D(z) = \int_0^\infty K_D(z,z') \,\delta \omega(z')\,dz',
\label{eq:distance_response_kernel}
\end{equation}
where the distance kernel takes the generic form
\begin{equation}
K_D(z,z') = -\frac{3c}{2} \frac{\Omega_{\rm DE}(z')}{1+z'} \int_{z'}^{z} \frac{dz_1}{H(z_1)} \,\Theta(z - z') \equiv \mathcal{K}_D(z,z') \, \Theta(z - z') \,,
\label{eq:distance_kernel_theta}
\end{equation}
with $\Omega_{\rm DE}(z)$ denoting the DE density fraction. Here $\Theta(z-z')$ is the Heaviside step function, and $\mathcal{K}_D(z,z')$ is a smooth function determined by nested integrals of the background expansion rate.

Equation~\eqref{eq:distance_kernel_theta} makes explicit that distance observables at redshift $z$ are sensitive only to the DE evolution at earlier redshifts $z' \le z$. This causal ordering reflects the fact that distances are obtained by integrating the expansion history along the past light cone. As a result, perturbations in $\omega(z')$ at different redshifts contribute cumulatively to $D(z)$ and can partially compensate each other. The Fisher matrix for a discretized representation of the EoS, $\omega(z) \rightarrow \{\omega_i\}$, is then given by
\begin{equation}
F_{ij}^{(D)} = \sum_{\alpha,\beta} \frac{\partial D_\alpha}{\partial \omega_i} \, (C^{-1})_{\alpha\beta} \, \frac{\partial D_\beta}{\partial \omega_j}, \label{eq:fisher_distance_only}
\end{equation}
where each derivative involves an integral of the kernel $\Theta(z_\alpha - z')\,\mathcal{K}_D(z_\alpha,z')$ over the corresponding redshift bin.

\subsection{Eigenvalue hierarchy and causal smoothing}

Diagonalizing the distance-only Fisher matrix according to Eq.~\eqref{eq:fisher_diagonalization}, we find a strongly hierarchical eigenvalue spectrum \begin{equation} \lambda_1 \gg \lambda_2 \ge \lambda_3 \ge \cdots . \end{equation}
This hierarchy arises because the Heaviside structure of the kernel enforces a strong smoothing of the DE evolution. Only broad, slowly varying combinations of $\omega(z)$ survive the cumulative integration, while modes that oscillate or vary rapidly in redshift are efficiently suppressed~\cite{Lee:2025jrr}.

Operationally, this implies that distance and expansion data constrain only a single dominant linear combination of the cosmological parameters
\begin{equation} v_1 = a\,\Omega_{m0} + b\,H_0 + c\,\omega_0 + d\,\omega_a + e\,\sigma_8, \end{equation} corresponding to the eigenvector associated with the largest eigenvalue $\lambda_1$ of the Fisher matrix. All orthogonal combinations are suppressed by much smaller eigenvalues and thus remain weakly constrained or effectively degenerate. This behavior has been noted in earlier analyses of distance-based DE constraints \cite{Maor:2000jy,Huterer:2002hy} and is independent of the specific parametrization adopted for the equation of state.

Importantly, this eigenvalue hierarchy is insensitive to the statistical precision of the measurements~\cite{Lee:2025jrr}. Improving distance accuracy rescales the overall normalization of the Fisher matrix but does not alter the step-function structure of the kernel. Consequently, the strong hierarchy persists even for high-precision distance and expansion data, providing a direct explanation for the intrinsic insensitivity of distance-based observables to multi-dimensional time dependence in the DE EoS. The Fisher eigenvalues discussed here should not be interpreted as forecasts of parameter uncertainties. Instead, their relative hierarchy diagnoses how many independent directions in parameter space are actually probed by a given set of observables. In particular, a strongly hierarchical spectrum with $\lambda_1 \gg \lambda_2$ indicates that the data effectively constrain only a single linear combination of parameters, regardless of the dimensionality of the model space. The analysis focuses on eigenvalue ratios and their evolution, rather than on absolute eigenvalue magnitudes.

In the next section, we show that growth observables possess a qualitatively different kernel structure, which lifts subdominant eigenmodes and enables sensitivity to additional, genuinely independent directions in parameter space.

\section{Impact of Growth Measurements}
\label{sec:Growth}

In this section, we demonstrate explicitly how the inclusion of growth observables modifies the structure of the Fisher information matrix and alleviates the strong eigenvalue hierarchy characteristic of distance-only analyses. The key result is that growth measurements introduce genuinely new information directions by virtue of their differential and partially local response to the DE EoS.

\subsection{Kernel structure revisited}

As shown in Appendix~\ref{app:distance_kernel}, the response of distance observables to a perturbation $\delta\omega(z)$ is governed by a kernel of the form
\begin{equation}
K_D(z,z') = \mathcal{K}_D(z,z')\,\Theta(z-z'),
\end{equation}
which enforces a strictly cumulative and causal dependence on the expansion history. This structure leads to strong smoothing and suppresses sensitivity to localized or rapid time variation.

In contrast, the growth factor $D_{+}(a)$ responds to $\delta\omega(z)$ through the linearized growth equation, whose solution admits the exact kernel decomposition
\begin{equation}
K_G(z,z') = \delta(z-z')\,\mathcal{K}_G^{\rm loc}(z) + \Theta(z'-z)\,\mathcal{K}_G^{\rm int}(z,z'),
\label{eq:growth_kernel_summary}
\end{equation}
as derived in Appendix~\ref{app:growth_kernel}. The presence of the Dirac-delta contribution reflects the explicit local dependence of the growth equation on derivatives of the background expansion rate. This term has no analogue in distance observables and plays a crucial role in lifting degeneracies. Physically, this local term originates from the explicit dependence of the growth equation on the derivative of the Hubble rate, and is therefore absent in purely geometric observables.

\subsection{Fisher matrix structure with growth}

When growth observables such as $f\sigma_8(z)$ are included in the data vector, the Fisher matrix acquires additional contributions
\begin{equation}
F_{ij} = F_{ij}^{(D+H)} + F_{ij}^{(G)},
\end{equation}
where $F_{ij}^{(D+H)}$ denotes the distance and expansion-rate component discussed in Sec.~\ref{sec:Hierarchy}, and $F_{ij}^{(G)}$ arises from growth measurements. Because $F_{ij}^{(G)}$ inherits both local and integrated kernel components from Eq.~\eqref{eq:growth_kernel_summary}, it is not aligned with the dominant eigenmode of $F_{ij}^{(D+H)}$. Instead, growth observables probe parameter combinations that are nearly orthogonal to the distance-dominated direction in parameter space. As a result, subdominant eigenvalues of the total Fisher matrix are significantly enhanced.

Diagonalizing the combined Fisher matrix, we find that the eigenvalue spectrum becomes substantially less hierarchical
\begin{equation}
\lambda_1 > \lambda_2 \sim \mathcal{O}(0.1\,\lambda_1),
\end{equation}
with additional modes gaining non-negligible constraining power. This behavior signals the emergence of a second independent information direction, corresponding to genuine sensitivity to the time dependence of the DE EoS.

\subsection{Physical interpretation}

The physical origin of this qualitative change is straightforward. Distance observables compress the expansion history into a single accumulated measure, so that different histories with similar integrals remain observationally degenerate. Growth observables, by contrast, respond to the instantaneous competition between Hubble friction and gravitational clustering. A perturbation in $\omega(z)$ therefore leaves a localized imprint on the growth rate at nearby redshifts, which cannot be fully mimicked by compensating changes at other epochs. This enhancement of sensitivity is structural rather than statistical. It does not rely on unrealistically precise growth data or on specific DE parametrizations. Instead, it follows directly from the differential nature of the growth equation and the resulting kernel structure. Even moderate-precision growth measurements are sufficient to activate additional Fisher eigenmodes once they reach a threshold where the local kernel contribution becomes comparable to the cumulative distance term.

\subsection{Consequences for dynamical dark energy tests}

The analysis presented here clarifies why claims of time-varying DE based solely on distance data are often unstable or prior dependent. In the absence of growth information, the Fisher matrix effectively constrains only a single averaged mode of $\omega(z)$, regardless of how many parameters are introduced. The inclusion of growth observables fundamentally changes this situation by increasing the dimensionality of the information accessible to the data.

In the following section, we discuss the implications of this result for current and future surveys, with particular emphasis on how survey configurations that combine geometric and growth measurements can robustly test DDE beyond the limitations of distance-only analyses.

\section{Implications for Survey-like Data}
\label{sec:Implications}

The results of Secs.~\ref{sec:Hierarchy} and \ref{sec:Growth} establish a structural understanding of why distance-based probes alone are intrinsically limited in their ability to test time-dependent DE, and how growth information can qualitatively alter the accessible parameter space. In this section, we make these statements quantitative by explicitly examining the Fisher eigenvalue spectra obtained in representative survey-like setups.

Rather than focusing on forecasted parameter uncertainties, we emphasize the structure of the Fisher information matrix itself. In particular, we track how the inclusion of additional observables modifies the hierarchy of Fisher eigenvalues and activates genuinely new, independent Fisher eigenmodes. Throughout this section, numerical results are interpreted at the level of the eigenvalue spectrum; for compactness, we also report the effective information dimension $d_{\rm eff}$, defined from the normalized Fisher eigenvalues as a summary measure of how many eigenmodes contribute appreciably to the total information content. Denoting by $\{\lambda_i\}$ the eigenvalues of the Fisher matrix $\mathbf{F}$, we define the effective information dimension using the standard participation-ratio form
\begin{equation}
d_{\rm eff} \equiv \frac{\left(\sum_i \lambda_i\right)^2}{\sum_i \lambda_i^2}.
\label{eq:deff_pr_app}
\end{equation}
This quantity provides a compact summary of the Fisher eigenvalue spectrum and measures how many eigenmodes carry non-negligible weight. From this perspective, the number of meaningfully constrained parameter combinations is determined not by the dimensionality of the parameter space itself, but by how the information is distributed among the Fisher eigenvalues. This viewpoint underlies many discussions of parameter degeneracies and effective constraints in cosmological inference \cite{Huterer:2002hy,Raveri:2018wln}. Importantly, $d_{\rm eff}$ is used here only as a diagnostic indicator and does not replace the eigenmode-based interpretation.

We consider two complementary cases: controlled mock data experiments designed to isolate the effect of adding growth observables, and an Euclid-like configuration that reflects the precision and redshift coverage of next-generation surveys.

\begin{figure}[t]
\centering
\begin{minipage}{0.48\linewidth}
\centering
\includegraphics[width=\linewidth]{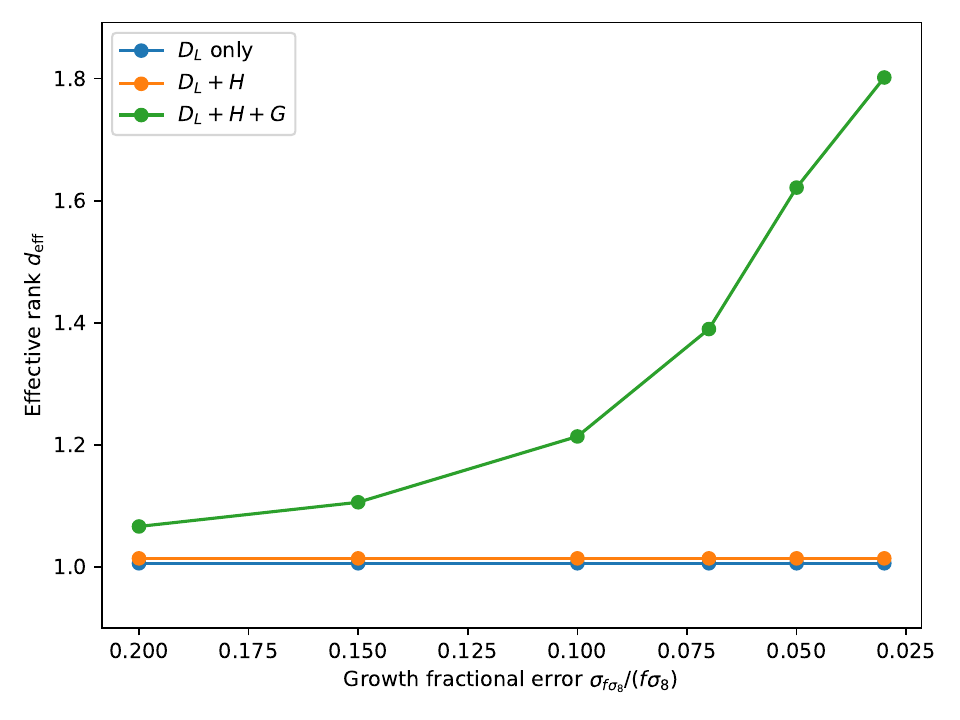}
\par\vspace{1mm}
\end{minipage}\hfill
\begin{minipage}{0.48\linewidth}
\centering
\includegraphics[width=\linewidth]{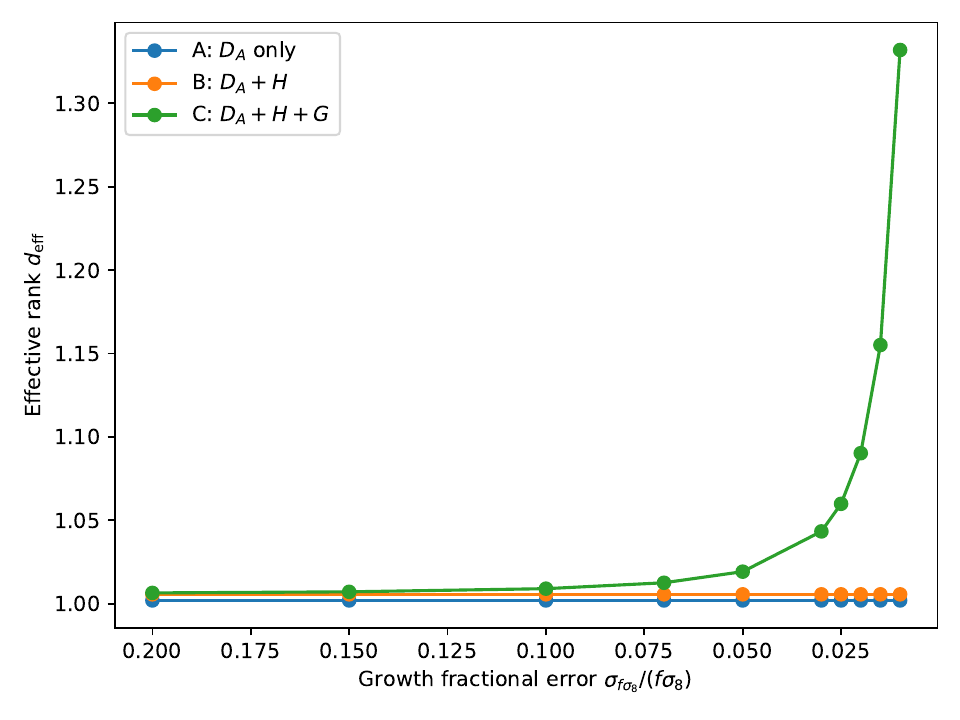}
\par\vspace{1mm}
\end{minipage}
\caption{Evolution of the Fisher eigenvalue hierarchy as a function of the fractional uncertainty of growth measurements, $\sigma_{f\sigma_8}$, with all other observational assumptions held fixed. The left panel shows results from controlled mock data experiments, while the right panel corresponds to Euclid-like survey specifications. In both cases, distance-only observables produce a strongly hierarchical Fisher spectrum dominated by a single eigenmode. When growth information is added, subdominant eigenvalues are lifted, indicating the emergence of additional information directions. In the controlled mock configuration (left), the second Fisher eigenmode becomes partially active already at moderate growth precision, leading to a gradual increase in the effective information dimensionality. In contrast, the Euclid-like configuration (right) exhibits a pronounced plateau, with the information remaining effectively one-dimensional down to the few-percent level. Only once the growth uncertainty reaches the percent or sub-percent regime does a rapid transition occur, signaling the activation of a second, genuinely independent Fisher eigenmode.}
\label{fig:eigen_scan}
\end{figure}

\subsection{Controlled mock data experiments}
\label{subsec:mock}

We begin with controlled mock data experiments combining distance, expansion rate, and growth observables. The mock setup is deliberately simplified to make the kernel-driven origin of the information hierarchy transparent. Distance and expansion rate measurements probe the background expansion through cumulative line-of-sight integrals, while growth observables respond through a differential equation and therefore encode more localized information in redshift.

To make the numerical setup explicit, Table~\ref{tab:mock_vs_euclid_numeric} summarizes the concrete choices adopted in the controlled mock data analysis and contrasts them with the Euclid-like configuration discussed in Sec.~\ref{subsec:euclidlike}. The table lists the redshift sampling, fractional error models, covariance assumptions, and growth-precision ranges used in the two cases. All theoretical ingredients---including the fiducial cosmology, parameter set, finite-difference scheme, and numerical solvers---are held fixed. The controlled mock configuration is therefore distinguished solely by its simplified and uniform observational assumptions, designed to isolate the intrinsic information structure of the observables independently of survey-specific details.

\begin{table}[t]
\centering
\caption{Numerical configuration used in the two deterministic Fisher scripts. The theory model, growth ODE solver, parameter set, and finite-difference scheme are identical; only the data configuration (redshift sampling, error model, and covariance structure) differs.}
\label{tab:mock_vs_euclid_numeric}
\begin{tabular}{lcc}
\hline\hline
Item & Controlled mock & Euclid-like  \\
\hline
Distance observable & $D_L(z)$ & $D_A(z)$ \\
Redshift sampling (distance) & $z\in[0.1,1.2]$, $n_{DL}=20$ (uniform) & $z=\{0.9,1.0,\dots,1.8\}$, $n=10$ \\
Redshift sampling ($H$) & $z\in[0.1,1.2]$, $n_{H}=15$ (uniform) & same $z$ bins as $D_A$ \\
Redshift sampling (growth) & $z\in[0.1,1.2]$, $n_{G}=15$ (uniform) & same $z$ bins as $D_A$ \\
Fractional errors (distance) & $\sigma_{D_L}/D_L = 0.02$ (2\%) & $\sigma_{D_A}/D_A = 0.005$ (0.5\%) \\
Fractional errors ($H$) & $\sigma_{H}/H = 0.03$ (3\%) & $\sigma_{H}/H = 0.007$ (0.7\%) \\
Fractional errors (growth baseline) & $\sigma_{f\sigma_8}/(f\sigma_8)=0.05$ (5\%) & $\sigma_{f\sigma_8}/(f\sigma_8)=0.05$ (5\%) \\
Growth error scan list & \multicolumn{2}{c}{$\{0.20,\,0.15,\,0.10,\,0.07,\,0.05,\,0.03,\,0.025,\,0.02,\,0.015,\,0.01\}$} \\
Covariance structure & diagonal only & per-bin $(D_A,H)$ block with $\rho_{D_AH}=0.4$ \\
Growth solver & \multicolumn{2}{c}{RK4 in $\ln a$, $a_{\rm ini}=10^{-3}$, $N_{\rm step}=6000$} \\
Finite-difference steps & \multicolumn{2}{c}{$(\Delta\Omega_{m0},\Delta H_0,\Delta \omega_0,\Delta \omega_a,\Delta\sigma_8)=(10^{-4},10^{-2},10^{-4},10^{-4},10^{-3})$} \\
Fiducial cosmology & \multicolumn{2}{c}{$(\Omega_{m0},H_0,\omega_0,\omega_a,\sigma_8)=(0.30,\,70,\,-1,\,0,\,0.80)$} \\
\hline\hline
\end{tabular}
\end{table}

For distance observables alone (Case~A), the Fisher eigenvalue spectrum is strongly hierarchical
\begin{align}
\lambda_i^{(A)} = \big( 1.13\times10^{4},\; 3.34\times10^{1},\; 4.07\times10^{-1},\; 1.26\times10^{-2},\; 0
\big).
\end{align}
We emphasize that the absolute magnitude of the Fisher eigenvalues depends on the normalization of the observables and the  assumed covariance. Only their relative hierarchy and ratios carry physical meaning in the present analysis. The exactly vanishing eigenvalue reflects the absence of sensitivity to the growth normalization $\sigma_8$. More generally, the extreme hierarchy $\lambda_1 \gg \lambda_{i>1}$ indicates that essentially all accessible information is concentrated in a single dominant Fisher eigenmode, consistent with the causal smoothing discussed in Sec.~\ref{sec:Hierarchy}.

Including expansion-rate measurements (Case~B) increases the overall Fisher normalization and partially lifts subdominant eigenvalues
\begin{align}
\lambda_i^{(B)} = \big( 2.44\times10^{4},\; 1.70\times10^{2},\; 1.14,\; 9.26\times10^{-2},\; 0 \big),
\end{align}
but the Fisher spectrum remains strongly hierarchical. This is reflected in the fact that the information continues to be dominated by the leading eigenmode, with $d_{\rm eff}^{(B)}\simeq1.01$ serving only as a compact summary of this dominance. Therefore, expansion rate data sharpen the existing geometric mode but do not activate a new independent Fisher eigenmode.

A qualitative transition occurs only once growth information is included (Case~C). In this case, the Fisher spectrum becomes
\begin{align}
\lambda_i^{(C)} = \big( 2.82\times10^{4},\; 9.28\times10^{3},\; 3.27\times10^{2},\; 2.70,\; 3.12\times10^{-1} \big),
\end{align}
indicating that multiple eigenvalues now carry non-negligible weight. Equivalently, the second Fisher eigenmode becomes dynamically relevant, and the information is no longer confined to a single direction. This change is summarized by $d_{\rm eff}^{(C)}\simeq1.6$, but its physical origin lies in the activation of additional eigenmodes rather than in the numerical value of this scalar diagnostic.

The left panel of Fig.~\ref{fig:eigen_scan} illustrates how this transition depends on the precision of growth data. As the fractional uncertainty in growth measurements decreases, the ratio $\lambda_2/\lambda_1$ increases steadily, marking the emergence of a second independent Fisher eigenmode.  The mild plateau observed in the controlled mock case reflects the residual dominance of correlated geometric information, which suppresses sensitivity to time-dependent modes until the local growth kernel becomes sufficiently strong. Although the numerical results shown here are obtained using a specific parametrization of the equation of state, the emergence of a second Fisher eigenmode can be traced to the kernel structure of growth observables and is therefore not tied to any particular parametrization choice.

It is important to emphasize that the early activation of subdominant Fisher eigenmodes observed in the controlled mock experiments should not be interpreted as a generic feature of growth measurements. Rather, it reflects the deliberately simplified structure of the mock data, including uniform redshift sampling, diagonal covariance, and minimal overlap between geometric and growth information. Under these conditions, the growth kernel is only weakly suppressed by geometric degeneracies, allowing a second information direction to become partially accessible even at the $\sim10\%$ level in growth precision.

As we show in Sec.~\ref{subsec:euclidlike}, this behavior changes qualitatively once realistic survey features are introduced. In Euclid-like configurations, correlations between distance and expansion observables, together with survey-specific binning and covariance structure, maintain a strong coupling between geometry and growth. As a result, the effective information dimensionality remains pinned near $d_{\rm eff}\simeq1$ over an extended range of growth precision, producing a pronounced plateau that persists down to the few-percent level. The contrast between the controlled mock and Euclid-like cases demonstrates that the emergence of additional Fisher eigenmodes is not governed by growth precision alone, but by how effectively growth information decouples from geometric degeneracies within the full data covariance.

\subsection{Euclid-like growth information and threshold behavior}
\label{subsec:euclidlike}

We next consider an Euclid-like configuration to assess whether the same eigenvalue-level transition persists under more realistic survey conditions. The theoretical framework and Fisher construction are identical to those used in the controlled mock data analysis; only the redshift sampling, error amplitudes, and covariance structure are modified.

For completeness, the explicit Fisher eigenvalues $\lambda_i^{(A,B,C)}$ obtained in the Euclid-like configuration at the reference growth precision $\sigma_{f\sigma_8}=0.05$ are reported in Table~\ref{tab:euclid_eigenvalues}. As emphasized throughout this work, these eigenvalues serve primarily as diagnostic quantities, while their relative hierarchy and evolution with growth precision carry the physical content of interest.

\begin{table}[t]
\centering
\caption{Fisher eigenvalues and effective information dimension for the Euclid-like configuration at the reference growth precision $\sigma_{f\sigma_8}=0.05$.}
\label{tab:euclid_eigenvalues}
\begin{tabular}{lcccccc}
\hline\hline
Case & $d_{\rm eff}$ & $\lambda_1$ & $\lambda_2$ & $\lambda_3$ & $\lambda_4$ & $\lambda_5$ \\
\hline
A: $D_A$ only
& $1.00194$
& $2.16\times10^{5}$
& $2.09\times10^{2}$
& $7.24\times10^{-2}$
& $6.14\times10^{-5}$
& $0$ \\

B: $D_A+H$
& $1.00559$
& $9.27\times10^{5}$
& $2.59\times10^{3}$
& $5.58$
& $2.00\times10^{-2}$
& $0$ \\

C: $D_A+H+G$
& $1.01919$
& $9.28\times10^{5}$
& $6.26\times10^{3}$
& $2.62\times10^{3}$
& $6.18$
& $2.74\times10^{-2}$ \\
\hline\hline
\end{tabular}
\end{table}

For Euclid-like angular-diameter distances alone, the Fisher matrix remains strongly dominated by a single eigenmode, with $d_{\rm eff}^{(A)}\simeq1.00$. Adding expansion-rate measurements increases the Fisher trace but leaves the eigenvalue hierarchy essentially unchanged, with $d_{\rm eff}^{(B)}\simeq1.01$. Even at high precision, distance and expansion-rate observables therefore remain confined to a single dominant Fisher eigenmode.

Including growth information leads to a modest but genuine change in the eigenvalue structure. The second eigenvalue $\lambda_2$ becomes non-negligible compared to $\lambda_1$, signaling the onset of a second, growth-driven Fisher eigenmode. For the reference growth precision $\sigma_{f\sigma_8}=0.05$, this behavior is summarized by $d_{\rm eff}^{(C)}\simeq1.02$. As the growth uncertainty is reduced, however, the contribution of the second eigenmode increases rapidly, leading to a corresponding rise in $d_{\rm eff}$. As in the mock data case, the essential point is therefore not the absolute value of the summary statistic at a fixed precision, but the emergence and growth of an additional independent Fisher eigenmode.

An important feature of the Euclid-like results is the rapid rise of the second eigenvalue once the fractional uncertainty of growth measurements approaches the $\sim2\%$ (percent-level) regime.  In particular, when the growth precision reaches the $\sim 2\%$ regime, the ratio $\lambda_2/\lambda_1$ increases significantly, signaling the activation of a second independent Fisher eigenmode.

This threshold behavior is particularly evident in the right panel of Fig.~\ref{fig:eigen_scan}. In contrast to the controlled mock-data case, the Euclid-like configuration exhibits an extended plateau in which $d_{\rm eff}$ remains close to unity over a wide range of growth precisions. This reflects the fact that, under realistic survey conditions, residual parameter degeneracies and correlated distance--expansion information continue to suppress sensitivity to time-dependent dark energy until growth measurements reach sufficiently high precision. Once the percent-level threshold is crossed, however, the second Fisher eigenmode is rapidly activated, leading to a sharp increase in $\lambda_2/\lambda_1$ and a correspondingly abrupt rise in $d_{\rm eff}$. The right panel highlights that the emergence of genuinely new information directions is not gradual, but occurs only when growth data become precise enough to overcome the intrinsic smoothing imposed by geometric observables.

This transition has a clear physical interpretation. The newly activated eigenmode corresponds to sensitivity to genuinely time-dependent components of DE, which are strongly suppressed in distance-based observables by cumulative line-of-sight integration. Percent-level growth measurements therefore open an information direction that is structurally inaccessible to purely geometric probes.

Importantly, this does not imply that any specific parametrization of DDE is tightly constrained. Rather, it indicates a change in the information geometry itself: once growth precision crosses the percent-level threshold, variations in the temporal behavior of DE project onto a distinct Fisher eigenmode that was previously unobservable.

Taken together, the controlled mock data and Euclid-like analyses demonstrate that improved distance precision alone cannot overcome the intrinsic smoothing imposed by distance kernels. Only observables with localized, differential sensitivity—such as growth measurements—can activate additional Fisher eigenmodes and enable meaningful tests of DDE.

\section{Discussion}
\label{sec:Dis}

In this work, we have examined the fundamental capability of cosmological observables to probe time variation in the dark energy equation of state from an information-structural perspective. Rather than focusing on parameter constraints within a specific model or survey, our analysis isolates how different classes of observables populate the Fisher information matrix and how many independent directions in parameter space are actually accessible to the data.

A central result is that distance-based observables, including luminosity and angular-diameter distances as well as direct measurements of the expansion rate, are intrinsically limited to a strongly hierarchical Fisher spectrum. The dominant eigenvalue overwhelmingly exceeds all subdominant modes, reflecting the cumulative and causal structure of the distance kernels. As a consequence, distance-only data effectively constrain a single linear combination of cosmological parameters, even when the underlying model space is multidimensional. This limitation is structural rather than statistical: improving measurement precision rescales the overall Fisher normalization but does not alter the eigenvalue hierarchy.

The inclusion of growth observables qualitatively changes this situation. Because growth responds through a differential equation rather than an integrated kernel, it introduces localized sensitivity to the expansion history. At the level of the Fisher matrix, this manifests as the lifting of subdominant eigenvalues and the emergence of an additional independent information direction. The transition is not gradual but exhibits threshold-like behavior: below a certain growth precision, the information remains effectively one-dimensional, whereas beyond that threshold, the second eigenmode becomes observationally relevant. This behavior provides a concrete eigenvalue-level criterion for when genuinely time-dependent aspects of DE can be tested.

An important implication is that apparent evidence for DDE derived from distance data alone should be interpreted with caution. Since only one effective direction is constrained, multi-parameter reconstructions of $\omega(z)$ are necessarily prior-dependent and unstable. In this sense, the present analysis helps clarify why different parameterizations or basis choices often lead to qualitatively different conclusions when applied to distance-only data. The issue does not originate from the choice of parametrization itself, but from the limited dimensionality of the information encoded in the observables.

Our use of mock data and Euclid-like survey configurations serves to demonstrate that these conclusions are not artifacts of idealized assumptions. The mock data experiments isolate the kernel-driven information hierarchy in a controlled setting, while the Euclid-like configuration shows that next-generation growth measurements operate in the regime where additional Fisher directions are activated.  The same theoretical framework and numerical machinery are used in both cases, ensuring that the observed differences arise from the observables rather than from methodological choices.

Several caveats are worth emphasizing. First, the Fisher analysis employed here is deterministic and based on local derivatives around a fiducial cosmology. While this is sufficient for diagnosing information dimensionality, it is not intended as a replacement for full likelihood or MCMC analyses when precise parameter inference is the goal. Second, our treatment assumes Gaussian errors and simplified covariance structures; incorporating more realistic systematics may shift the quantitative thresholds but is unlikely to change the qualitative conclusions regarding the eigenvalue hierarchy.

More broadly, the information-based viewpoint adopted here suggests a reframing of how future survey results should be interpreted. Rather than asking only whether parameter uncertainties shrink, it is equally important to assess whether new observables activate genuinely independent information directions. From this perspective, growth measurements play a qualitatively distinct role, not merely tightening constraints but expanding the dimensionality of what can be learned about cosmic acceleration.

In summary, our results provide a structural explanation for the long-standing difficulty of testing DDE with distance data alone and identify growth observables as a necessary ingredient for overcoming this limitation. By focusing on the eigenvalue structure of the Fisher matrix, we offer a parametrization-independent criterion for assessing when claims of dynamical dark energy are physically meaningful.

\section{Conclusions}
\label{sec:Con}

In this work, we have investigated the fundamental capability of cosmological observables to test dynamical dark energy from an information-theoretic perspective. By analyzing the eigenvalue structure of the Fisher information matrix, we have shown that distance-based probes—including luminosity and angular-diameter distances, as well as direct measurements of the expansion rate—are intrinsically limited to a strongly hierarchical information spectrum. As a consequence, such observables effectively constrain only a single dominant combination of cosmological parameters, even when the underlying model space allows for multiple degrees of freedom. This limitation arises from the cumulative and causal structure of the distance kernels and cannot be overcome by improvements in measurement precision alone.

The inclusion of growth observables leads to a qualitative change in the accessible information structure. Because growth responds through differential dynamics rather than integrated kernels, it can in principle activate additional Fisher eigenmodes that are structurally inaccessible to purely geometric measurements. At sufficiently high precision, a second independent information direction becomes observable, providing genuine sensitivity to time-dependent components of the dark-energy equation of state. Crucially, this transition is encoded in the hierarchy of Fisher eigenvalues and can be identified independently of any specific parametrization of dark energy.

Our numerical experiments, based on both controlled mock data analyses and Euclid-like survey configurations, demonstrate that this behavior is robust and not an artifact of idealized assumptions. The same theoretical framework, fiducial cosmology, and numerical machinery are applied throughout, ensuring that the observed differences arise from the intrinsic properties of the observables and their covariance structure rather than from methodological choices.

A central result of this study is the existence of a precision-dependent transition in the effective information dimensionality. In controlled mock configurations, where the data are intentionally simplified (e.g.\ uniform redshift sampling, minimal covariance, and weak overlap between geometry and growth information), growth measurements with fractional uncertainties at the $\sim10\%$ level already partially activate a second Fisher eigenmode, raising the effective information dimension to $d_{\rm eff}\gtrsim1$. In this case, the contribution of growth information is only weakly suppressed, and the accessible information dimension increases gradually with improving precision.

In contrast, realistic Euclid-like survey configurations exhibit a pronounced plateau, with $d_{\rm eff}\simeq1$ persisting down to the $\sim2$--$3\%$ level in growth precision. Here, correlations among distance and expansion observables, together with the survey’s binning and covariance structure, maintain a strong degeneracy between geometry and growth information. As a result, even when growth data are included, the inference remains confined to a single dominant Fisher eigenmode until growth measurements reach percent-level or sub-percent precision. Only below this threshold does a second eigenmode emerge rapidly, enabling robust sensitivity to temporal variations in dark energy.

These results clarify that the often-invoked few-percent criterion for testing dynamical dark energy is not universal, but applies specifically to realistic survey-like configurations. More generally, they highlight that the impact of growth observables depends not only on their nominal precision, but also on how effectively they decouple from geometric information within the full covariance structure of the data.

From a broader perspective, our findings suggest a shift in emphasis for future studies of cosmic acceleration. Beyond reporting increasingly tight parameter constraints, it is essential to assess whether new data genuinely expand the dimensionality of the accessible information in parameter space. From this viewpoint, growth observables play a qualitatively distinct role: they are not merely a means of refining constraints along already-dominant directions, but a mechanism for opening new information directions when survey design and precision are sufficient.

In conclusion, distance-based observations alone are structurally incapable of testing multi-dimensional time dependence in the dark-energy equation of state. Growth measurements are a necessary ingredient for overcoming this limitation, but their effectiveness is conditional. Only when measurement precision and survey configuration allow growth information to decouple from geometric degeneracies does the effective information dimension exceed unity. This information-theoretic perspective provides a constructive guideline for the design and interpretation of future surveys aimed at probing the dynamical nature of cosmic acceleration.

\section*{Acknowledgments}

This work is supported by the Basic Science Research Program through the National Research Foundation of Korea (NRF), funded by the Ministry of Science and ICT under Grant Nos.~NRF-RS-2021-NR059413 and NRF-2022R1A2C1005050.

\appendix
\section{Response Kernels for Distance and Growth Observables}
\label{app:kernels}

In this Appendix, we provide the explicit derivation of the kernel structures used in the main text to describe the response of distance and growth observables to perturbations in the dark energy equation of state. The results presented here underlie Eqs.~\eqref{eq:distance_response_kernel} and~\eqref{eq:distance_kernel_theta} and clarify the origin of the causal and local contributions discussed in Secs.~\ref{sec:Hierarchy} and~\ref{sec:Growth}.

\subsection{Distance kernel}
\label{app:distance_kernel}

The comoving radial distance is given by
\begin{equation}
\chi(z) = \int_0^z \frac{c\,dz_1}{H(z_1)} ,
\end{equation}
so that its variation under a perturbation $\delta\omega(z)$ of the dark energy equation of state is
\begin{equation}
\delta\chi(z) = -\int_0^z \frac{c\,dz_1}{H(z_1)}\, \frac{\delta H(z_1)}{H(z_1)} .
\label{eq:app_dchi}
\end{equation}

From the Friedmann equation and the dark energy continuity equation, the
fractional variation of the Hubble rate can be written exactly as
\begin{equation}
\frac{\delta H(z_1)}{H(z_1)} = \frac{3}{2} \int_0^{z_1} \frac{\Omega_{\rm DE}(z_2)}{1+z_2} \, \delta\omega(z_2)\,dz_2 ,
\label{eq:app_dH_over_H}
\end{equation}
where $\Omega_{\rm DE}(z)$ is the dark energy density fraction.

Substituting Eq.~\eqref{eq:app_dH_over_H} into Eq.~\eqref{eq:app_dchi} and
exchanging the order of integration yields
\begin{equation}
\delta\chi(z) = \int_0^z dz_2\, \left[ -\frac{3c}{2}\, \frac{\Omega_{\rm DE}(z_2)}{1+z_2} \int_{z_2}^{z}\frac{dz_1}{H(z_1)} \right]
\delta\omega(z_2) .
\end{equation}

This expression defines the distance kernel
\begin{equation}
\delta\chi(z) = \int_0^\infty K_\chi(z,z')\,\delta\omega(z')\,dz',
\end{equation}
with
\begin{equation}
K_\chi(z,z') = \Theta(z-z') \left[ -\frac{3c}{2}\, \frac{\Omega_{\rm DE}(z')}{1+z'} \int_{z'}^{z}\frac{dz_1}{H(z_1)} \right].
\label{eq:app_distance_kernel}
\end{equation}

This kernel exhibits a purely integrated (Heaviside) structure, reflecting the cumulative nature of distance observables. The same background response relations derived here will be used in the analysis of growth observables (Appendix~\ref{app:growth_kernel}).
\subsection{Growth kernel}
\label{app:growth_kernel}

In this Appendix, we derive the kernel representation for the response of the linear growth factor $D_{+}$ to $\delta\omega(z)$. Unlike distance observables, growth responds through a second-order differential equation, leading to both localized and integrated contributions in the response kernel. Throughout this derivation, we make use of the background response relations for $\delta\ln H$ and $\delta\Omega_m$ derived in Appendix~\ref{app:distance_kernel}, and focus here on the additional structures that arise from the differential nature of the growth equation.

\paragraph{Background relations.}

We use $x\equiv\ln a$ as the time variable and define $E(a)\equiv H(a)/H_0$. The functional derivative of the background expansion with respect to $\omega(x')$ follows directly from Eq.~\eqref{eq:app_dH_over_H} and can be written compactly as
\begin{equation}
\frac{\delta\ln H(x)}{\delta\omega(x')} = -\frac{3}{2}\,\Omega_{\rm DE}(x)\,\Theta(x-x') . \label{eq:app_dlnH_dw}
\end{equation}

Differentiating with respect to $x$ yields
\begin{equation}
\frac{\delta}{\delta\omega(x')} \left(\frac{d\ln H}{dx}\right) = -\frac{3}{2} \left[ \frac{d\Omega_{\rm DE}(x)}{dx}\,\Theta(x-x')
+ \Omega_{\rm DE}(x)\,\delta(x-x') \right], \label{eq:app_dlnHprime_dw}
\end{equation}
where the Dirac-delta term arises from differentiating the step function.

Similarly, the variation of the matter fraction is
\begin{equation}
\frac{\delta\Omega_m(x)}{\delta\omega(x')} = 3\,\Omega_m(x)\Omega_{\rm DE}(x)\,\Theta(x-x') .
\label{eq:app_dOmegam_dw}
\end{equation}

\paragraph{Linearized growth equation.}
The exact growth equation is
\begin{equation}
D_{+}''(x) + A(x)\,D_{+}'(x) - \frac{3}{2}\Omega_m(x)D_{+}(x) = 0, \qquad A(x)\equiv 2+\frac{d\ln H}{dx},
\label{eq:app_growth_ode}
\end{equation}
where primes denote derivatives with respect to $x$. Varying Eq.~\eqref{eq:app_growth_ode} with respect to $\omega(x')$ yields the exact inhomogeneous equation for $\delta D_{+}(x)$
\begin{equation}
\mathcal{L}\bigl[\delta D_{+}(x)\bigr] = -\delta A(x)\,D_{+}'(x) +\frac{3}{2}\,\delta\Omega_m(x)\,D_{+}(x),
\label{eq:app_var_growth}
\end{equation}
with the linear operator
\begin{equation}
\mathcal{L}[y]\equiv y''+A(x)y'-\frac{3}{2}\Omega_m(x)y.
\end{equation}
Using Eqs.~\eqref{eq:app_dlnHprime_dw} and~\eqref{eq:app_dOmegam_dw}, the functional derivatives of the coefficient variations are
\begin{align}
\frac{\delta A(x)}{\delta\omega(x')} &= -\frac{3}{2}\left[ \frac{d\Omega_{\rm DE}(x)}{dx}\,\Theta(x-x') + \Omega_{\rm DE}(x)\,\delta(x-x') \right], \label{eq:app_dA_dw} \\
\frac{\delta\Omega_m(x)}{\delta\omega(x')} &= 3\,\Omega_m(x)\Omega_{\rm DE}(x)\,\Theta(x-x').
\label{eq:app_dOm_dw_again}
\end{align}

\paragraph{Green-function solution and kernel decomposition.}
Let $D_{+}(x)$ be the growing-mode solution of the homogeneous equation $\mathcal{L}[D_{+}]=0$, and let $D_{-}(x)$ be an independent decaying-mode solution. Define the Wronskian
\begin{equation}
W(x)\equiv D_{+}(x)D_{-}'(x)-D_{-}(x)D_{+}'(x).
\end{equation}
The retarded Green function for $\mathcal{L}$ is then given exactly by
\begin{equation}
G(x,s) = \Theta(x-s)\, \frac{D_{+}(x)D_{-}(s)-D_{-}(x)D_{+}(s)}{W(s)}.
\label{eq:app_green}
\end{equation}
Assuming the normalization of $D_{+}$ is fixed at an early epoch $x_i$ and that $\delta D_{+}(x_i)=\delta D_{+}'(x_i)=0$, the solution of Eq.~\eqref{eq:app_var_growth} can be written as
\begin{equation}
\delta D_{+}(x) = \int_{x_i}^{x} ds\, G(x,s)\, \left[-\delta A(s)\,D_{+}'(s) +\frac{3}{2}\delta\Omega_m(s)\,D_{+}(s) \right].
\label{eq:app_dD_solution}
\end{equation}

We now express $\delta D_{+}(x)$ in kernel form with respect to $\delta\omega(x')$
\begin{equation}
\delta D_{+}(x) = \int dx'\,K_G(x,x')\,\delta\omega(x').
\end{equation}
Using Eqs.~\eqref{eq:app_dA_dw}--\eqref{eq:app_dOm_dw_again} in Eq.~\eqref{eq:app_dD_solution}, one obtains the exact decomposition
\begin{equation}
K_G(x,x') = \delta(x-x')\,\mathcal{K}_G^{\rm loc}(x) + \Theta(x-x')\,\mathcal{K}_G^{\rm int}(x,x'),
\label{eq:app_growth_kernel_final}
\end{equation}
where the localized and integrated components are
\begin{align}
\mathcal{K}_G^{\rm loc}(x) &= +\frac{3}{2}\,G(x,x)\,\Omega_{\rm DE}(x)\,D_{+}'(x), \label{eq:app_KG_loc} \\
\mathcal{K}_G^{\rm int}(x,x') &= \int_{x'}^{x} ds\,G(x,s)\, \Biggl\{ -\frac{3}{2}\left[\frac{d\Omega_{\rm DE}(s)}{ds}\right]D_{+}'(s)
+\frac{9}{2}\,\Omega_m(s)\Omega_{\rm DE}(s)\,D_{+}(s) \Biggr\}. \label{eq:app_KG_int}
\end{align}
Equations~\eqref{eq:app_growth_kernel_final}--\eqref{eq:app_KG_int} provide the explicit forms of the growth kernels used in the main text.

\paragraph{Conversion to redshift variables.}
If one prefers kernels in $z$ rather than $x$, note that $dx = d\ln a = -dz/(1+z)$ and
\begin{equation}
\delta(x-x') = (1+z)\,\delta(z-z'), \qquad \Theta(x-x')=\Theta(z'-z),
\end{equation}
so the decomposition in Eq.~\eqref{eq:app_growth_kernel_final} translates straightforwardly to the redshift domain with the appropriate Jacobian factors.

\paragraph{Growth observables.}
Finally, the growth rate is defined exactly as $f=d\ln D_{+}/d\ln a = D_{+}'/D_{+}$, and $\sigma_8(z)=\sigma_8\,D_{+}(z)/D_{+}(0)$. Therefore, the response of $f\sigma_8(z)$ to $\delta\omega(z')$ inherits the same kernel structure, with additional (exact) factors from the chain rule.



\end{document}